\documentclass[prl, showpacs, preprint, floatfix]{revtex4}
\usepackage{latexsym}
\usepackage{amssymb}
\usepackage{amsmath}\large\huge\normalsize
\usepackage{graphicx}
\usepackage{amsfonts}
\usepackage{array}
\usepackage{subfigure}

\newcommand{\be}{\begin{equation}}
\newcommand{\ee}{\end{equation}}
\newcommand{\bea}{\begin{eqnarray}}
\newcommand{\eea}{\end{eqnarray}}

\begin{document}
\title{Quantum Corrected Spherical Collapse: A Phenomenological Framework}

\author{Jonathan Ziprick}
\email[Electronic address:]{jziprick@perimeterinstitute.ca}
\affiliation{Perimeter Institute and Department of Physics and Astronomy, University of Waterloo, Waterloo, Ontario, CANADA, N2L 3G1}

\author{Gabor Kunstatter}
\email[Electronic address:]{g.kunstatter@Uwinnipeg.ca}
\affiliation{Department of Physics and Winnipeg Institute of Theoretical Physics, University of Winnipeg, Winnipeg, Manitoba, CANADA, R3B 2E9}

%\affiliation{Department of Physics and Winnipeg Institute of Theoretical Physics, ${}^*$University of Manitoba and ${}^\dagger$University of Winnipeg, Winnipeg, Manitoba, CANADA}

\received{\today}

%\vspace{0.5in}

\begin{abstract}
A phenomenological framework is presented for incorporating quantum gravity motivated corrections into the dynamics of spherically symmetric collapse. The effective equations are derived from a variational principle that guarantees energy conservation and the existence of a Birkhoff theorem. The gravitational potential can be chosen as a function of the areal radius to yield specific non-singular static spherically symmetric solutions that generically have two horizons.
For a specific choice of potential the effective stress energy tensor violates only the dominant energy condition. The violations are maximum near the inner horizon and die off rapidly.
A numerical study of the quantum corrected collapse of a spherically symmetric scalar field in this case reveals that 
the modified gravitational potential prevents the formation of a central singularity and ultimately yields a static, mostly vacuum, spacetime with two horizons. The matter "piles up" on the inner horizon giving rise to mass inflation at late times. The Cauchy horizon
is transformed into a null, weak singularity, but in contrast to Einstein gravity, the absence of a central singularity renders this null singularity stable.
\end{abstract}


\maketitle
%\noindent
\section{Introduction}
%{\it Introduction.} 

%%%Spherically symmetric collapse of the thin null shell of mass M is considered in the framework of a local theory describing vacuum polarization effects--Frolov Vilkovisky 1981

Einstein's theory predicts its own demise in the form of cosmological and black hole singularities. One of the goals of quantum gravity is to provide a mathematically consistent mechanism for avoiding these singularities.  Recently, minisuperspace models derived from loop quantum gravity have successfully resolved singularities in quantum cosmology \cite{lqc} and black hole interiors \cite{Bojowald, Modesto, pullin, peltola09}. Nonetheless, in the absence of a complete microscopic quantum description of black holes, it is useful to take a  phenomenological (or in the more descriptive language of reference \cite{barrabes} "zoological") approach in which one tries to catalogue the possible non-singular semi-classical black hole spacetimes that can in principle emerge from the microscopic theory. This approach has a fairly long and distinguished history with much work focusing on the construction of non-singular static black hole spacetimes. Such metrics are presumed to be solutions to quantum corrected Einstein equations with an effective stress energy tensor on the right hand side. Bardeen \cite{bardeen} was one of the first to write down such a non-singular static black hole, but many other spacetimes have since been constructed \cite{poisson1,mars,peltola09}. These solutions generically have a pair of horizons, with the inner horizon radius determined by the dimensionful parameter associated with the short distance behaviour. The conditions under which such black holes can exist were clarified in the mid-nineties by Borde \cite{borde} in response to the proposal in \cite{mars} for a non-singular Schwarzschild black hole that satisfied the weak energy condition. Inner horizons are associated with mass inflation \cite{mass inflation} and one may ask whether the resulting spacetimes are non-singular and stable under perturbations (i.e. tossing in more matter). 

Other methods also exist for constructing non-singular static black holes.  In \cite{frolov} for example, the Schwarzschild metric inside a black hole was matched to a deSitter interior along a spacelike surface which was interpreted as a thin transition layer between the classical and quantum regions.  Alternatively one can solve semi-classical equations of motion in a quantized mini-superspace model as done recently in spherically symmetric loop quantum gravity \cite{Bojowald,Modesto,pullin}.

A potentially more fruitful approach is to examine singularity resolution in the context of dynamical black hole spacetimes. One of the first models  was proposed many years ago by Frolov and Vilkovisky \cite{frolov_vilkovisky81} (see \cite{roman83} for an interesting subsequent analysis). In this model a trapping surface forms but the inner horizon does not reach $r=0$. In this case two final outcomes are possible: the apparent horizons stabilize to yield a black hole with a static outer horizon, or the apparent horizons meet to form a closed trapping surface. In the latter case there is no true event horizon, although if the outer horizon is stable for long enough it will for all practical purposes mimic an event horizon over the relevant time scale. This scenario has recently been advocated by Hayward \cite{hayward}, and explicitly realized \cite{ziprick2} in a numerical calculation of spherically symmetric scalar field collapse using flat slice, or Painleve-Gullstrand (PG) coordinates (see \cite{ziprick1} for the corresponding purely classical numerical analysis in flat slice coordinates).  In this model, loop quantum gravity motivated corrections to the gravitational potential were incorporated directly into the scalar field equations of motion. The effect of these corrections on Choptuik scaling were analyzed, confirming the existence of a mass gap in the quantum corrected case, as observed by Husain \cite{husain1}. The choice of PG coordinates allowed the equations to be evolved beyond initial horizon formation thereby mapping out a large portion of the resulting spacetime, including the formation of trapping surfaces. For macroscopic black hole formation, two interesting features emerged: first the inner horizon was repelled by the repulsive core and no singularity formed; secondly, the modified equations of motion were non-conservative which allowed the outer horizon to shrink and the inner horizon to grow until they annihilated to form a closed trapping region. 

The purpose of the present paper is twofold. First, we present a systematic, general method for studying the formation of non-singular static black holes using dynamical equations that are derived from a variational principle.  A free function in the lagrangian allows us to specify the effective short distance behaviour of the gravitational potential which in turn determines the form of the vacuum solution. We look at a particular potential chosen to yield the Poisson-Israel \cite{poisson1} metric as a static solution. A straightforward calculation of the effective stress energy tensor reveals that the weak energy condition is generically satisfied, but the dominant energy condition (DEC) is violated outside the inner horizon. The violations die of asymptotically so that they are negligable near the outer horizon of a macroscopic black hole. For microscopic black holes these violations persist at the outer horizon and beyond. Since the violations of the DEC allow matter to propagate outside the light cone, this could contribute to the evaporation of microscopic black holes by increasing the rate of Hawking radiation or providing a new mechanism for mass loss.
% it yields the intriguing possibility for a phenomenological description of Hawking radiation.

Second, we give the results of a numerical study of the gravitational collapse of a massless scalar field using the same choice of potential as described above. The central singularity is indeed avoided. Despite violations of the DEC no energy escapes the outer horizon and there is no mass loss, thus two static horizons eventually form. The matter ``piles up" at the inner horizon, and as expected from previous studies \cite{mass inflation} of mass inflation, the mass function grows exponentially along this (nearly) null surface.  The interior near the origin is flat, and the exterior spacetime settles down to the non-singular vacuum solution everywhere except along the inner horizon. 

The paper is organized as follows. In Section 2 we review several static, vacuum spacetimes that describe non-singular black holes, including the Bardeen metric and the Poisson-Israel metric. Section 3 describes the general action principle that we use to modify the gravitational potential.  We write down the static non-singular vacuum solutions and derive the effective stress energy tensor. We also present the dynamical equations for spherically symmetric scalar field collapse in PG coordinates. The gravitational self-interaction of the scalar field is specified in terms of a single free function (the quantum corrected graviational potential) of the areal radius $r$. Section 4 reviews the numerical method for evolving the dynamical equations while Section 5 presents the results. We conclude with a summary and some speculation.

%\label{sec: eqmo}
%\noindent
\section{Static Non-singular Black Holes}
%We assume that such solutions can generically be put in Schwarzschild form:
%\begin{equation}
%ds_{(4)}^2= -\left(1-\frac{2M}{j(r)}\right)dt^2+\left(1-\frac{2M}{j(r)}\right)^{-1}dr^2+r^2 d\Omega^2 \ .
%\label{nonsingular metric}
%\end{equation}
%This is still somewhat restrictive since one can in principle also consider:
%\begin{equation}
%ds^2_{(4)}= -\left(1-\frac{2M}{j(r)}\right)dt^2+\left(1-\frac{2M}{j(r)}\right)^{-1}dr^2+h(r) d\Omega^2 \ ,
%\label{general 4d}
%\end{equation}
%which is the case, for example, in the quantum corrected black hole studied in \cite{peltola09}. For simplicity we henceforth consider the simpler ansatz, but there are no impediments to extending the formalism, as will be discussed at the end.

We assume that such solutions can generically be put in the form:
\begin{equation}
ds^2_{(4)}= -\left(1-\frac{G^{(4)}M}{lj(r)}\right)dt^2+\left(1-\frac{G^{(4)}M}{lj(r)}\right)^{-1}dr^2+h(r) l^2 d\Omega^2 \ ,
\label{general 4d}
\end{equation}
which is the case, for example, in the quantum corrected black hole studied in \cite{peltola09}. Here $l$ is a parameter with dimensions of length, usually taken to be the Planck length, introduced for dimensional considerations.

In general, the function $j(r)$ will contain a parameter $\mu$ of dimension length and should go to $r$ as $r \to \infty$. Moreover, if the core is to be regular $1/j(r)$ should vanish at least as fast as $r^2$ as $r \to 0$, in which case the geometry near the origin will be de-Sitter like. With these conditions we see that $1/j(r)$ goes to $0$ at $r=0$ and $r=\infty$ so that there will generically be either no horizons, or two horizons, with a mass gap determined by the shape of $j(r)$ and the value of $\mu$.
For example, the Bardeen black hole \cite{bardeen} has:
\begin{equation}
j(r)= \frac{(r^2+\mu^2)^{3/2}}{2lr^2} \ .
\end{equation}
On the other hand Poisson and Israel \cite{poisson1} used semi-classical %quantum gravity arguments
considerations to argue for a non-singular static metric with:
\begin{equation}
j(r) = \frac{r^3+\mu^3}{2lr^2} \ .
\label{poisson israel}
\end{equation}
Their argument suggested the identification: $\mu^3=a^2 M_{adm}$, where $a^2$ is the quantum cut-off scale and $M_{adm}$ is the ADM mass. This had the advantage that the size of the inner horizon did not go to zero as the black hole mass increased. 

In the following we will present a lagrangian formalism for deriving gravitational equations that produce solutions with arbitrary (but fixed) $j$ and allows coupling to matter so that one can examine the %question of
dynamical formation of non-singular black holes. 

\section{Lagrangian Formulation and Equations of Motion}

Our starting point is the action for generic dilaton gravity in two dimensional spacetime:
\be 
S[g,\phi ] = \frac{1}{2G}\int d^2x  \sqrt{-g}\, \bigg( \phi R(g) + S_M[g,\phi,\psi]
	+ \frac{V(\phi)}{l^2}\bigg),
\label{2d action}
\ee
where $G$ is the 2-d gravitational constant.
% and $l$ is a parameter of unit length, generally taken to be the Planck length.
The dilaton $\phi$ and metric $g$ are geometrical variables and $S_M$ is a matter action that will be specified below. The most general solution can be written in the form:

\be \label{eq:ds}
ds^2 = -j(\phi)\left[ \left( 1-\frac{2lG \cal{M}}{j(\phi)}\right) dt^2 +\left(1-\frac{2lG \cal{M}}{j(\phi)}\right)^{-1}dr^2 \right] ,
\ee
where ${\cal M} \equiv {\cal M}(t,r)$ is a generalized Misner-Sharpe mass function that is related to the traditional definition \cite{misner sharpe} by ${\cal M}_{\hbox{\scriptsize trad}}  = r {\cal M} / j(\phi)$. The vacuum theory with a constant mass $M$ satisfies a Birkhoff theorem \cite{birkhoff}. $j(\phi )$ is related to the dilaton potential by:
\be \label{eq:j} 
\frac{dj}{d\phi}=V(\phi ) \ ,
\ee
and $r=r(\phi)$ is determined by:
\be
dr=l\frac{d\phi}{j(\phi)} \ .
\ee
This action describes a wide class of theories containing black hole solutions, including the spherically symmetric sector of D-dimensional Einstein gravity. The latter requires the identifications:
\bea
2G&=& \frac{16\pi G^{(n+2)} n}{8(n-1){\cal \nu}^{(n)}l^n},\\
\phi&=&\frac{n}{8(n-1)}\left(\frac{r}{l}\right)^n,\\
\label{eq:correspondence3} V(\phi)&=& (n-1)\left(\frac{n}{8(n-1)}\right)^{1/n}\phi^{-1/n}, \\
h(\phi)&=& \frac{8(n-1)}{n}\phi=\left(\frac{r}{l}\right)^n,
\eea
where $G^{(n+2)}$ is the $D$-dimensional Newton's constant, $r$ is the radius of a rotational invariant two-sphere, and
\be {\cal \nu}^{(n)}=\frac{2\pi^{(n+1)/2}}{\Gamma (\frac{1}{2}(n+1))}
\ee
is the volume of the $n$-dimensional unit sphere. The physical D-dimensional metric is:
\be
ds_{(D)}^2= \frac{1}{j(\phi)}ds^2+r^2(\phi)d\Omega_{(n)}^2 \ ,
\label{4d metric}
\ee
where $d\Omega_{(n)}^2$ is the volume element of the unit $n$-sphere. We have chosen to use $h(r)$ as for D-dimensional Einstein gravity, though one is free to choose a different form.
It is straightforward to verify that with the above identifications, the following $D=n+2$ dimensional metric is the correspondng Tangerlini-Schwarzschild solution:
\be
ds_{(D)}^2= -\left(1-\frac{2l^{D-2}GM}{r^{D-3}}\right)dt^2 +\left(1-\frac{2l^{D-2}GM}{r^{D-3}}\right)^{-1}dr^2 + r^2 d\Omega_{(n)}^2 \ .
\label{tangerlini}
\ee
In the following we will focus on $n=2$, i.e. 4-dimensional spherically symmetric %Einstein
gravity. %, for which $j(r)=j(\phi(r))=r$.
The generalization to higher dimensions will be straightforward. 

We propose that the action (\ref{2d action}) be used to generate the dynamical equations of motion for spherically symmetric collapse in 4-dimensions with quantum corrected gravitational potential. In particular, one defines the physical 4-d metric in terms of 2-d quantities by (\ref{4d metric}), first choosing the function $j(r)$ and then specifying $\phi(r)$ and $V(\phi)$ in the action to give this $j(r)$. That is, one requires:
\bea
ld\phi &=& j(r)dr \ ; \\
V(\phi)&=& \frac{dj(\phi)}{d\phi}=\frac{dj(r)}{dr}\frac{dr}{d\phi}=\frac{l}{j(r)}\frac{dj}{dr} \ .
\eea
For example, for the Poisson-Israel metric (\ref{poisson israel}), which will be the focus of our numerical studies, one obtains:
\bea
l\phi(r)&=&\int dr \frac{(r^3+\mu^3)}{2lr^2}=\frac{r^2}{4l}-\frac{\mu^3}{2lr} \ ;\\
V(\phi)&=& \frac{l(r^3-2\mu^3)}{r(r^3+\mu^3)} \ .
\eea
The above procedure yields a vacuum solution that corresponds to the static spherically symmetric 4-d metrics of precisely the form 
%(\ref{nonsingular metric}).
\ref{general 4d}).
Note that both $\phi$ and $V(\phi)$ go to minus infinity as $r\to0$. This behaviour is generic for any function with asymptotic form $j(r) \sim r^{-\alpha}$ with $\alpha \ge 1$. Only $j(r)$ enters the equations of motion in the subsequent analysis, so these divergences do not affect the resulting dynamics.

Once the gravitational part of the action has been chosen, one can study the quantum corrected dynamics of spherically symmetric collapse by specifying the matter lagrangian. For simplicity we consider a massless scalar field, but there is no obvious impediment to study other forms of matter, including electromagnetic and Yang-Mills fields. Our matter lagrangian is:
\be
L_M= \int d^2x r^2(x) \sqrt{-g}\left|\nabla\psi\right|^2 \ ,
\ee
where we have made explicit the fact that $r(x)$ is a scalar function of the coordinates, $\nabla$ refers to a 2-d gradient, and we have been consistent in using $h(r) = r^2$ in this lagrangian as well.
%Had we chosen to generalize the metric to the form (\ref{general 4d}) 4-d general covariance would dictate that $r^2$ be replaced by $h(r)$ in the matter action as well.

\subsection{Effective Stress Energy Tensor}

Given the 4-d metric (\ref{4d metric}) it is straightforward to calculate the 4-d Einstein tensor and write the equations of motion in 4-d form.  The result is:
\bea
G^{(4)}_{\mu\nu}&=& 8\pi G^{(4)}\frac{r}{2lj(r)}\left(\nabla_\mu\psi\nabla_\nu\psi -\frac{1}{2}g^{(4)}_{\mu\nu}\left|\nabla\psi\right|^2\right)
    + T^{\hbox{\scriptsize eff}}_{\mu\nu} \ ; \\
   G^{(4)}_{ij}&=& -\frac{8\pi}{2} G^{(4)}\frac{r}{2lj(r)} \left|\nabla\psi\right|^2 + T^{\hbox{\scriptsize eff}}_{ij} \ .
\eea 
In the above $\mu,\nu =0,1$ refer to the spherically symmetric coordinates $(t,x)$,  while $i,j = 2,3$ refer to the angular coordinates. The modified action gives rise to two effects. First, there is an effective $r$-dependent gravitational constant:
\be
G^{\hbox{\scriptsize eff}}=G^{(4)}\frac{r}{2lj(r)}=G^{(4)}\frac{r^3}{r^3+\mu^3} \ ,
\ee
which vanishes as $r\to 0$. Secondly, there is an effective stress energy tensor  of the form:
\bea
T^{\hbox{\scriptsize eff}}_{\mu\nu}&=& \frac{2G^{\hbox{\scriptsize eff}}\cal M}{r}\left(\frac{j'}{j r}- \frac{1}{r^2}\right)g^{(4)}_{\mu\nu}=:\rho g^{(4)}_{\mu\nu} \ ,\\
T^{\hbox{\scriptsize eff}}_{ij}&=& \frac{2G^{\hbox{\scriptsize eff}}\cal M}{r} \left(\frac{j'}{j r}- \left(\frac{j'}{j}\right)^2+\frac{1}{2}\frac{j''}{j}\right)g^{(4)}_{ij}=:\beta g^{(4)}_{ij} \ ,
\eea
where the $'$ denotes differentiation with respect to $r$.
%and ${\cal M}$ is the generalized Misner-Sharpe mass function \cite{misner sharpe} defined by: THIS NEEDS SOME CHECKING: I AM NOT SURE HOW BEST TO DEFINE THIS.
%\be
%\frac{2G^{(4)}{\cal M}}{j(r)}= -\left|\nabla r\right|^2-1
%\label{mass function}
%\ee 
%where $-|\nabla r|^2$ gives the expansion of outgoing null geodesics, so that 
This effective stress tensor is only non-vanishing when the mass function %${\cal M}$
is non-zero.

The mass function determines the location of apparent horizons via:
\be
1-\frac{G^{(4)}{\cal M}}{j(r)}=0 \ .
\label{horizon}
\ee
%It is perhaps interesting to note that 
If one uses the traditional definition of the Misner-Sharpe mass function
%\be
%\frac{2G^{(4)}{\cal M}_{trad}}{r}= -\left|\nabla r\right|^2-1
%\label{traditional mass function}
%\ee
then the horizon location is given by:
\be
1-\frac{2G^{\hbox{\scriptsize eff}}{\cal M}_{\hbox{\scriptsize trad}}}{r}=0 \ .
\ee

\begin{figure}[htb]
\begin{center}
\includegraphics{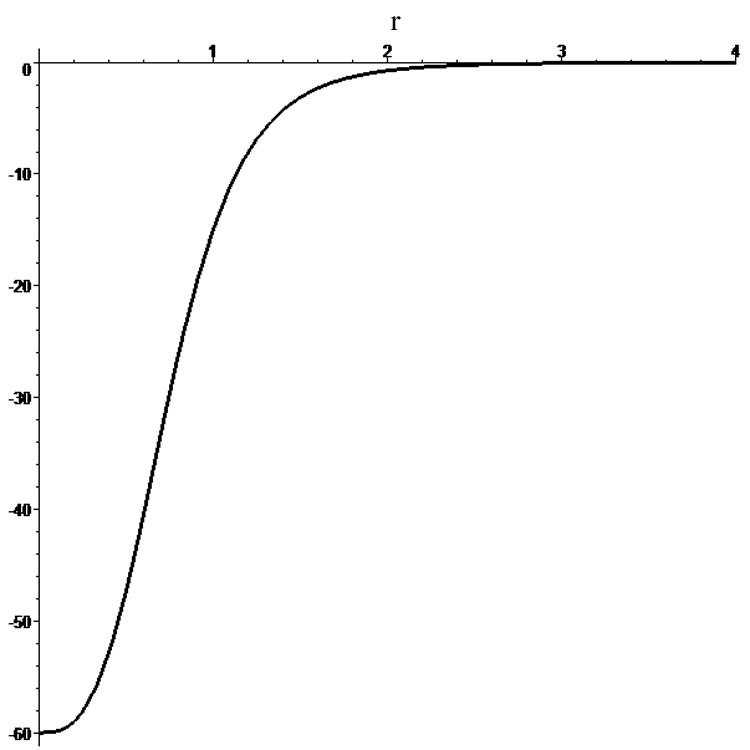}
\caption[]{A plot of the weak energy condition for $\mu=1$ and $M=5$.}
\label{WEC}
\end{center}
\end{figure}

\begin{figure}[htb]
\begin{center}
\includegraphics{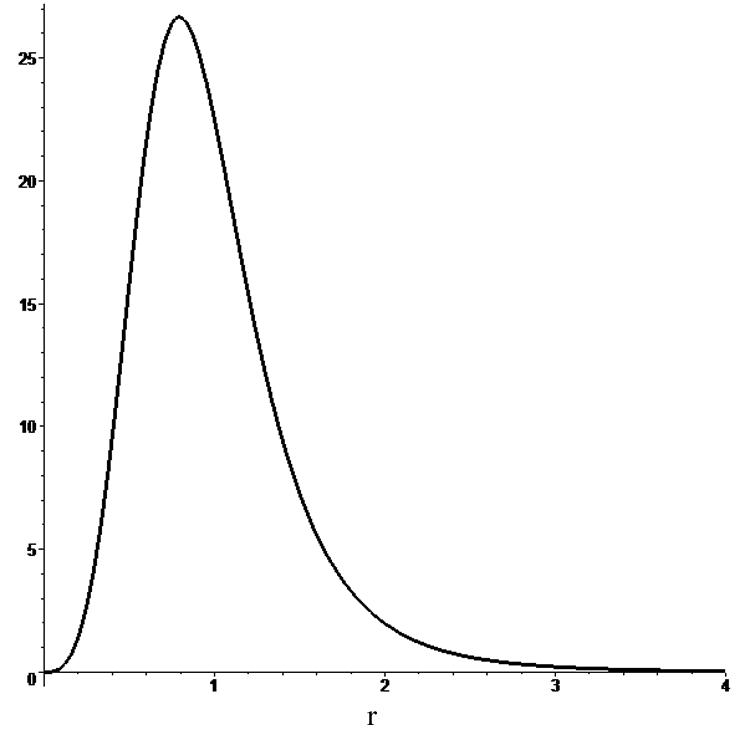}
\caption[]{A plot of the null and strong energy conditions for $\mu=1$ and $M=5$.}
\label{NEC}
\end{center}
\end{figure}

\begin{figure}[htb]
\begin{center}
\includegraphics{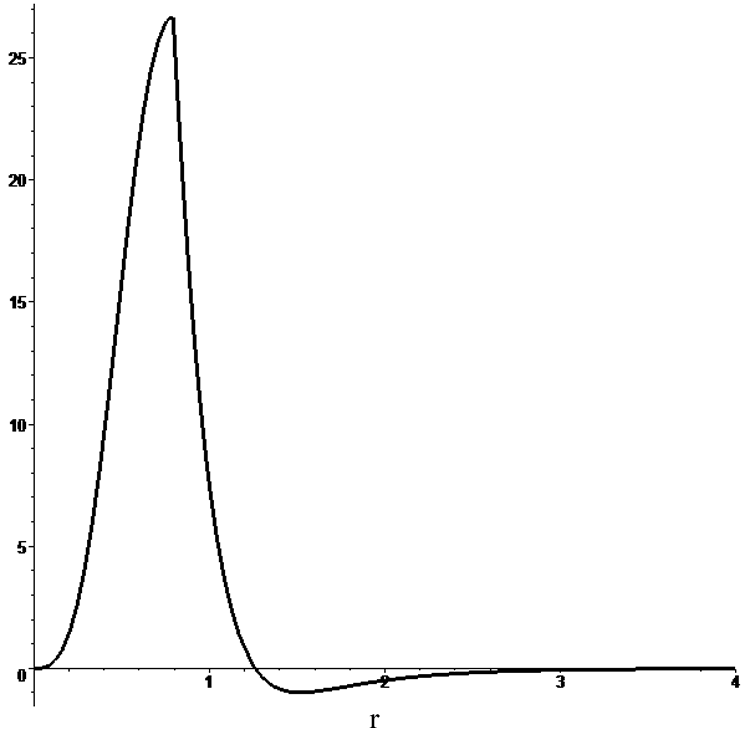}
\caption[]{A plot of the dominant energy conditon for $\mu=1$ and $M=5$. With these parameters, the inner horizon is located at $r_{ih} \approx 0.225$ while the outer horizon is at $r_{oh} \approx 10$. The curve crosses the axis at $r=2^{1/3}$.}
\label{DEC}
\end{center}
\end{figure}

Since $r/j(r)$ is a monotonic function for the class of theories we wish to consider, as long as the matter action is standard, violations of the energy conditions can only come from the effective stress tensor. Its simple form allows a detailed analysis. In fact, one can verify the following inequalities are required by the corresponding energy conditions:
\bea
\rho &<& 0 \qquad \hbox{Weak Energy Condition (WEC)} \ ;\nonumber\\
\beta -\rho &>& 0 \qquad \hbox{Null (NEC) and Strong (SEC) Energy Conditions} \ ;\nonumber \\
|\rho|-|\beta| &>& 0 \qquad \hbox{Dominant Energy Condition (DEC)} \ .
\eea
As shown in Figs. \ref{WEC}--\ref{DEC}, the WEC, NEC and SEC are satisfied for the $j(r)$ that we are using, but the DEC is violated outside the inner horizon. Notice that $1/j(r)$ has a positive slope for $r<2^{1/3}\mu$, which represents the extent of the repulsive core. Correspondingly, the DEC violations begin at this point and drop off rapidly as $r^{-6}$, so that they are not seen near the exterior of large black holes. However for microscopic black holes the violations can extend beyond the outer apparent horizon. Since the DEC dictates that mass-energy cannot flow faster than light, these violations could play a role in describing the evaporation of microscopic black holes within the theory.

Solving the horizon condition (\ref{horizon}) we find that the inner and outer horizons coincide at $r=2^{1/3} \mu$ giving the size of the smallest black hole allowed to form. This signifies a mass gap in the theory with a threshold mass of:
\begin{equation}
M_{th} = \frac{3 (2)^{1/3} \mu}{4G^{(4)}} \ .
\label{mthresh}
\end{equation}

\subsection{Hamiltonian Equations}
In order to look at spherically symmetric collapse and horizon formation it is convenient to use flat slice, or PG coordinates.
Choosing the space coordinate $x=r$, the generalized PG metric is given by
\begin{equation}
ds^2_{(4)} = -\sigma^2dt^2 + \left(dr + \sqrt{\frac{2lG{\cal M}}{j(r)}}\sigma dt\right)^2 + r^2 d\Omega^2 ,
\label{reduced metric}
\end{equation}
where $\sigma$ is the lapse function which scales time across a spatial slice. 
The gauge fixing procedure that determines the reduced Hamiltonian and equations of motion for the scalar field were derived in \cite{ziprick1}.
One first makes the gauge choice
\begin{equation}
\label{fix1}
j(\phi) = l \phi^\prime \,,
\end{equation}
which yields the consistency condition on the lapse and shift functions:
\begin{equation}
\label{sigma-N}
\sigma G \Pi_\rho = N \phi^\prime.
\end{equation}
The equations can be put in a more transparent form using the canonical transformation
\begin{eqnarray}
X&:=&e^\rho \ , \\
P&:=&e^{-\rho}G\Pi_\rho \ ,
\end{eqnarray}
where $\Pi_\rho$ is the momentum conjugate to $\rho$. $P$ is then conjugate to $X$ with their Poisson bracket being
\begin{equation}
\left\{X(x),P(y)\right\}=G\delta(x,y).
\end{equation}
The resulting Hamiltonian is
\begin{equation}
\begin{split}
H(X,P,\psi,\Pi_\psi)=&\int dr \; \sigma \left(-\frac{X^2}{j(\phi)}{\cal M}^\prime + {\cal G}_{\cal M} + l\frac{XP\psi^\prime \Pi_\psi}{j(\phi)}\right)\\
                     &+\int dr \left(\frac{\sigma X^2}{j(\phi)}{\cal M} \right)^\prime \ ,
\end{split}
\end{equation}
where
\begin{eqnarray}
{\cal M} &=& \frac{l}{2G}\left(P^2-\frac{(\phi^\prime)^2}{X^2}+\frac{j(\phi)}{l^2}\right) \ ,\label{eq:cal M}\\
{\cal G}_{\cal M} &=& \frac{1}{2}\left(\frac{\Pi_\psi^2}{h(\phi)} + h(\phi)(\psi^\prime)^2\right) \ .
\end{eqnarray}
${\cal M}$ is the Misner-Sharpe mass and approaches a constant at spatial infinity where it is equal to the ADM mass of the solution. ${\cal G}_{\cal M}$ is the energy density of the scalar field.

To completely fix the gauge we choose:
\begin{equation}
\label{fix2}
X=\sqrt{j(\phi)}.
\end{equation}
This condition, along with Eqs.(\ref{eq:cal M}) and (\ref{fix1}), implies that $P^2 = 2G{\cal M}/l$. These gauge conditions produce the non-static generalization of PG coordinates, as can be seen by using (\ref{sigma-N}) and the gauge conditions in the case of a vacuum to reduce the line element to PG form \cite{ziprick1}:
\begin{equation}
ds^2 = j(\phi)\left[-dt^2+\left(dr+\sqrt{\frac{2G{\cal M}l}{j(\phi)}}dt\right)^2\right].
\label{eq:vacuum PG}
\end{equation}
More importantly, the spatial slices are regular across apparent horizons that form during the evolution.

We are now able to write the equations of motion for the scalar field in fully reduced form:
\begin{eqnarray}
\label{psi dot}
\dot{\psi} &=& \sigma \left(\frac{l \sqrt{2G{\cal M}/l} \psi^\prime}{\sqrt{j(\phi)}} + \frac{\Pi_\psi}{h(\phi)}\right),\\
\label{Pi dot}
\dot{\Pi}_\psi &=& \left[\sigma \left( h(\phi)\psi^\prime + \frac{l \sqrt{2G{\cal M}/l} \Pi_\psi}{\sqrt{j(\phi)}} \right) \right]^\prime,
\end{eqnarray}
where $\sigma$ and ${\cal M}$ are the solutions to
\begin{equation}
\label{M eq}
{\cal M}^\prime =  {\cal G}_{\cal M}+l\psi^\prime \Pi_\psi \sqrt{\frac{2G{\cal M}l}{j(\phi)}},
\end{equation}
\begin{equation}
\label{sigma eq}
\sigma^\prime + \frac{G l \psi^\prime\Pi_\psi }{\sqrt{2G{\cal M}l j(\phi)} } \sigma = 0.
\end{equation}
The right hand side of (\ref{M eq}) defines the instantaneous mass density of the configuration. It has the expected contribution from the energy density of the scalar and an additional coupling term that corresponds to the contribution from its self gravity. 
%For the spatial integrations, we begin with $P(0,t)~=~0$ and $\sigma(0,t)~=~1.0$ which implies a flat spacetime at the origin.
 
The above equations need to be supplemented by boundary conditions for $\cal M$ and $\sigma$. Without loss of generality we choose $\sigma=1$ at $r=0$. A change in this value corresponds to a trivial rescaling of the time coordinate. For future reference we note that this condition implies that PG time $t$ and proper time $s$ are equivalent at the origin. We also fix ${\cal M}=0$ at $r=0$, which guarantees that the metric is flat in the neighbourhood of the origin. One way to think about this is that with this identification we have eliminated the single gravitational degree of freedom associated with the  spherically symmetric gravitational field \cite{kuchar} so that the gravitational mass is provided totally by the mass content of the spacetime.

\section{Numerical Methods}
%\noindent
%{\it Numerical Methods.}
The first step in the iteration process is to specify initial $\psi$ and $\Pi_\psi$ configurations. We work with two forms:
\begin{eqnarray}
\label{tanh}
\psi &=& A r^2 \exp{\left[-\left(\frac{r-r_0}{B}\right)^2\right]},\\
\label{gaussian}
\psi &=& A \tanh{\left(\frac{r-r_0}{B}\right)},
\end{eqnarray}
where $A$, $B$ and $r_0$ are the parameters which dictate the initial mass function. Note that since the mass density depends on $\psi^\prime$, the tanh form has one mass peak while the gaussian data has two. For both cases we define an initial standing wave by choosing $\Pi_\psi(r,0)=0$. The results were identical for both forms of initial data.

In order to have a fine grid spacing where the dynamics require a more accurate integration, we use a variable $r$-spacing that remains unchanged throughout the evolution. A typical $r$-lattice would use a spacing of $\Delta r(r)=10^{-3}$ from the origin until some point beyond the inner horizon, at which point the lattice increased by one percent each step until reaching a maximum of $10^{-2}$. With this maximum step, the lattice would be extended as far as necessary to contain all of the mass within the system (numerical errors occur when mass leaves the grid).

We use an adaptive time step $\Delta t(t)$ refinement according to the minimum found across the spacial slice with the condition
\begin{equation}
\Delta t(t) = \hbox{MIN}_r\left\{\frac{dt}{dr} \Delta r(r)\right\} ,
\end{equation}
where $\frac{dt}{dr}$ is the inverse of the local speed of an ingoing null geodesic. This provides stability by preventing information from moving over too many $r$-points in a single time step.

For derivatives, we experimented with finite differences and cubic splines finding the same qualitative behaviour using both, which gives some indication of reliability. In the end we chose to use finite differences exclusively since they require less computational time. The variable lattice was problematic for fourth order finite differences, so we used second order differences.

After specifying intial $\psi$ and $\Pi_\psi$, fourth order Runge-Kutta (RK4) methods are used to find $\cal M$ and $\sigma$ by integrating (\ref{M eq}) and (\ref{sigma eq}) across the spatial slice. These values are then entered into the right hand sides of (\ref{psi dot}) and (\ref{Pi dot}) to find $\psi$ and $\Pi_\psi$ at the next time step with RK4. This process is repeated until the matter disperses to leave behind a flat spacetime, or in the case of black hole formation, for as long as the code remained stable. In the latter case, the mass density was seen to bounce off the origin and move outward toward the inner horizon. As the mass moves closer and closer to the inner horizon, instability eventually sets in. However, the code always runs for long enough to extract the relevant information.

Convergence of our code is apparent since the results were not affected by increasing the resolution beyond $10^{-2}$ as long as the black hole size on formation was sufficiently large enough compared to the grid size. To ensure the stability of our code, we monitor the ADM mass for each run and find it to remain constant within $\sim 0.1\%$.

%\\[2pt]

\section{Results}
With the quantum scale parameter $\mu$ set to zero we find results identical to those in \cite{ziprick1}. If the mass profile on the initial time slice evolves such that ${\cal M}(r=2^{1/3} \mu) < M_{th}$ at all times, then no horizons form and the mass is free to disperse to infinity leaving behind a flat spacetime. If the threshold mass is reached, black hole formation begins with the appearance of a single horizon that splits into two as the evolution continues. The outer horizon continues to grow as mass falls through it while the inner horizon moves toward the origin.  The horizon paths define a two dimensional trapping surface in spacetime within which all null geodesics point toward the origin, illustrating several features expected in theory \cite{hayward}. Specifically the trapping horizon allows for multiple apparent horizons during intermediate phases of the collapse which annihilate in pairs. It necessarily terminates at the singularity at one end, and at null infinity (the horizon) at the other.

\begin{figure}[htb]
\begin{center}
\includegraphics{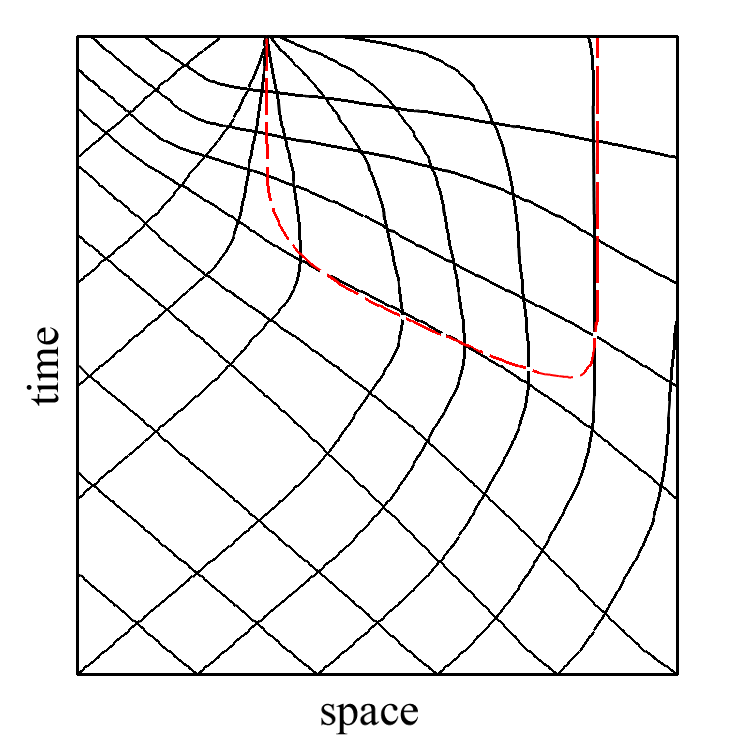}
\caption[]{A space-time diagram showing the trapping surface (red, dashed line) and null geodesic (black, solid lines). Notice all ``outward'' moving null geodesics inside the black hole converge at the inner horizon.}
\label{rt}
\end{center}
\end{figure}

For non-zero $\mu$ we find the mass of the smallest black hole to be $0.945 \mu/G^{(4)}$ in good agreement with the analytic calculation for static black holes (\ref{mthresh}). The outer horizon dynamics are similar to the classical case, however the repulsive core causes the inner horizon to slow as it moves toward $r \sim \mu$, eventually settling down to a constant value after all of the mass is inside the inner horizon. The mass continues moving inward toward the origin, but is now free to move through the origin and back out toward the inner horizon. As this mass nears the inner horizon, it continually slows so that it does not enter the trapping region in finite proper time. See Fig. \ref{rt} for a spacetime diagram showing the horizon locations and null geodesics throughout the evolution \footnote{An animation of the same space time showing the evolution of the mass density can be found at http://theoryX5.uwinnipeg.ca/users/jziprick/ (file named "quantum corrected collapse"). Note that the ordinate rescales throughout the evolution to keep the plot within view.}.

To investigate mass inflation, we solve for outgoing null geodesics by integrating: 
\begin{equation}
0 = \sigma dt - \left(1-\sqrt{\frac{2lG {\cal M}}{j}} \right)^{-1} dr \;
\end{equation}
We then define a coordinate $v$ along each outgoing null geodesic by:
\begin{equation}
dv = \sigma dt + \left(1+\sqrt{\frac{2lG {\cal M}}{j}} \right)^{-1} dr \ .
\end{equation}
Note that since we are not including the appropriate integrating factor the above does not define the coordinate transformation to Vaidya time. It does, however, define an affine parameter along each outgoing null geodesic and is sufficient to examine the issue of mass inflation. In particular, previous studies of classical black holes possessing an inner horizon (i.e. charged or rotating black holes) have found that along outgoing null geodesics originating near the origin, the mass function increased exponentially with Vaidya time, forming a weak \footnote{weak in the sense that the metric remains regular in suitable coordinates.} mass inflation singularity near the Cauchy horizon \cite{mass inflation, hod, ori}. In these scenarios the Cauchy horizon eventually contracts to the origin and forms a strong, spacelike singularity.

\begin{figure}[htb]
\begin{center}
\includegraphics{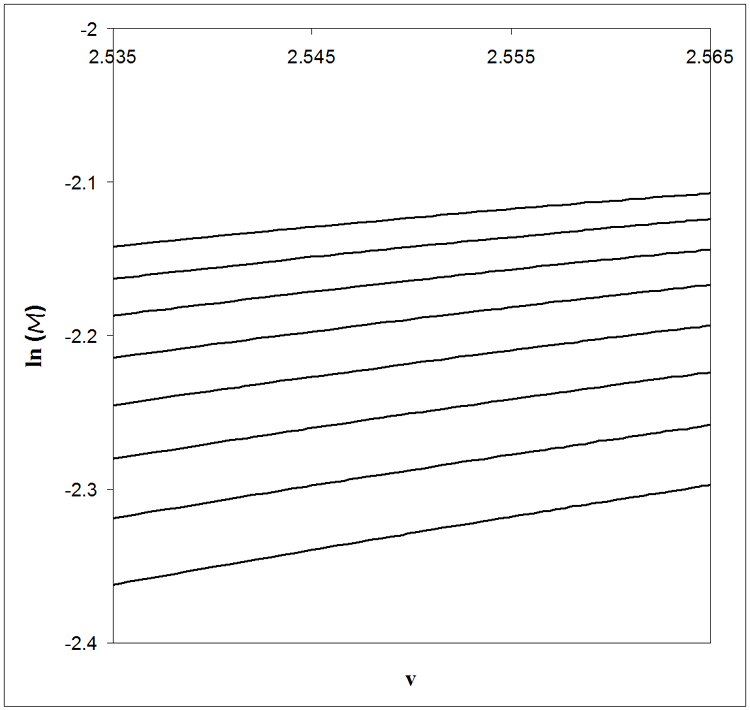}
\caption[]{Plot of $\ln(M)$ vs. $v$ for a series of null geodesics leaving the origin after black hole formation.}
\label{lnM}
\end{center}
\end{figure}

The plots of $\ln({\cal M})$ vs. $v$ in Fig. \ref{lnM} confirm mass inflation within the present quantum model. However in this case the inner horizon settles down to a constant radius, thereby removing the central strong singularity. A drawback of this model is that the final location of the inner horizon tends to zero for fixed $\mu$ with increasing mass, inevitabley forcing the mass toward the origin for macroscopic black holes. 

Note that the dynamical equations violate the DEC similarily to the vacuum solution, however this was not seen to allow for any black hole mass loss.

\section{Conclusion}
We have developed a Hamiltonian formulation of spherically symmetric gravitational collapse in the presence of quantum corrections. The form of the correction is freely specifiable and provides a computational laboratory for observing gravitational collapse in the presence of effective potentials arising from modifications to general relativity. 

With the choice of $j(r)$ studied herein, evolution of the dynamical equations results in either a flat spacetime or a black hole possessing two horizons, with the mass piling up along the inner horizon and forming a weak, null singularity. This family of corrections also permits a stable, vacuum solution. The collapse scenario and the static solution possess an equivalent mass gap that is proportional to the quantum scale parameter.

The modified potential was found to violate the DEC, which can allow for mass energy to flow faster than light. However, this effect was not observed in the numerical evolution; mass that entered the black hole was found to be trapped inside the inner horizon. A mass inflation singularity was found to form along the inner horizon. This horizon was stationary at late times rather than contracting to the origin as in classical Kerr or Reissner-Nordstr\"om black holes.

%However, we may speculate that a unitary evolution of this system is possible if the mass were to pass the inner horizon after $v \to \infty$. The resulting spacetime would be similar to the Reissner-Nordstr\"om case of infinitely oscillating universes, except without a central singularity.

An improvement to this model would be to consider some form of $j(r)$ that scales the quantum parameter $\mu$ so that the inner horizon does not go to zero for macroscopic black holes, similarly to the static solution studied by Israel and Poisson. Also, it would be quite interesting to incorporate a radiation term in the Hamiltonian to allow for evaporation which would likely lead to a closed trapping region as in \cite{hayward, ziprick2}. In this scenario one wonders how mass inflation would play out as the black hole shrinks and eventually vanishes.
 
\begin{acknowledgments}

{\bf Acknowledgements: }We thank D. Garfinkle, J. Gegenberg, V. Husain, H. Maeda, A. Peltola and E. Poisson for helpful discussions. G.K. is grateful to the Centro de Estudios Científicos, Chile for its kind hospitality. The authors are grateful to Westgrid for providing computer resources and the Natural Sciences and Engineering Research Council of Canada for financial support.
\end{acknowledgments}

\end{document}